\documentclass{article}
\usepackage[margin=1in]{geometry}
\usepackage{setspace}
\usepackage{cite}
\usepackage{caption}
\usepackage{amsmath, amssymb, amsfonts}
    
\usepackage{amsthm}
\usepackage{booktabs}  % for clean rules
\usepackage{wrapfig}   % for wraptable
\usepackage{amsmath}   % for math symbols
\usepackage{mathptmx} % Times font for text and math
\usepackage[cal=cm]{mathalpha} % Caligraphic fonts

\usepackage{algorithm}
\usepackage{algorithmicx}
\usepackage{algpseudocode}
\usepackage[font=small,labelfont=bf]{caption}

\algrenewcommand\algorithmiccomment[1]{\hfill\textcolor{gray}{\small\texttt{// #1}}}
\algrenewcommand\alglinenumber[1]{\small #1}
\algrenewcommand\textproc{} % Removes boldface in procedure calls

\usepackage{graphicx}
\usepackage{subcaption} % Modern alternative to subfig
\usepackage{float}

\usepackage{array}
\usepackage{tabularray}
\usepackage{makecell}
\usepackage{multirow}
\usepackage{diagbox}
\usepackage{booktabs}
\usepackage{threeparttable}

\usepackage{xcolor}
\usepackage{soul} % For highlighting
\usepackage{textcomp} % Additional text symbols

\usepackage{url}
\usepackage{hyperref} % Should be loaded last to prevent conflicts
\hypersetup{
    linkcolor=black,    % Internal links
}

\usepackage[switch]{lineno}

\usepackage{paralist} % For compact lists
\usepackage{tablefootnote}
\usepackage{pifont} % For checkmarks and x marks
\usepackage{balance} % To balance columns on the last page

\def\BibTeX{{\rm B\kern-.05em{\sc i\kern-.025em b}\kern-.08em
    T\kern-.1667em\lower.7ex\hbox{E}\kern-.125emX}}
\begin{document}
\title{Empirical Analysis of Asynchronous Federated Learning on Heterogeneous Devices: Efficiency, Fairness, and Privacy Trade-offs\thanks{This paper was accepted to IJCNN 2025. This version is a preprint and not the official published version.}
}
\date{}

\author{
  Samaneh Mohammadi\textsuperscript{1,2}\thanks{Corresponding author: \texttt{samaneh.mohammadi@ri.se}}, 
  Iraklis Symeonidis\textsuperscript{1}, 
  Ali Balador\textsuperscript{2}, 
  Francesco Flammini\textsuperscript{2} \\
  \textsuperscript{1}RISE Research Institutes of Sweden, Västerås, Sweden \\
  \textsuperscript{2}Mälardalen University, Västerås, Sweden
}

\maketitle

\begin{abstract}
Device heterogeneity poses major challenges in Federated Learning (FL), where resource-constrained clients slow down synchronous schemes that wait for all updates before aggregation. Asynchronous FL addresses this by incorporating updates as they arrive, substantially improving efficiency. While its efficiency gains are well recognized, its privacy costs remain largely unexplored—particularly for high-end devices that contribute updates more frequently, increasing their cumulative privacy exposure. This paper presents the first comprehensive analysis of the efficiency–fairness–privacy trade-off in synchronous vs. asynchronous FL under realistic device heterogeneity. We empirically compare \texttt{FedAvg} and staleness-aware \texttt{FedAsync} using a physical testbed of five edge devices spanning diverse hardware tiers, integrating Local Differential Privacy (LDP) and the Moments Accountant to quantify per-client privacy loss. Using Speech Emotion Recognition (SER) as a privacy-critical benchmark, we show that \texttt{FedAsync} achieves up to \textit{10×} faster convergence but exacerbates fairness and privacy disparities: high-end devices contribute \textit{6–10×} more updates and incur up to \textit{5×} higher privacy loss, while low-end devices suffer amplified accuracy degradation due to infrequent, stale, and noise-perturbed updates. These findings motivate the need for adaptive FL protocols that jointly optimize aggregation and privacy mechanisms based on client capacity and participation dynamics, moving beyond static, \textit{one-size-fits-all} solutions.
\end{abstract}
\vspace{1em}
\noindent\textbf{Keywords:} Federated Learning, Device Heterogeneity, Asynchronous FL, Differential Privacy.

\section{Introduction}
Federated Learning (FL)~\cite{mcmahan2017communication} enables edge devices to collaboratively train a shared model without sharing raw data, making it especially valuable for privacy-sensitive domains such as healthcare, finance, and IoT~\cite{kairouz2021advances, mohammadi2024balancing}. However, real-world FL deployments face critical challenges, particularly device heterogeneity, arising from variations in computational resources (e.g., CPU, GPU, RAM) across participating clients. This leads to the \textit{straggler effect}, where resource-limited devices slow down training in synchronous FL methods like \texttt{FedAvg}~\cite{mcmahan2017communication}, which require all client updates before aggregation.

To mitigate the straggler effect in heterogeneous environments, several strategies have been proposed~\cite{nishio2019client, nguyen2022federated, chai2020tifl,xie2019asynchronous}. 
Among these, asynchronous FL (e.g., \texttt{FedAsync}~\cite{xie2019asynchronous}) has emerged as one of the most effective, enabling immediate aggregation upon receiving a client update. This approach improves efficiency and incorporates staleness-aware weighting to reduce the instability caused by delayed contributions.

However, these efficiency gains may come with hidden costs. Clients with more powerful hardware contribute more frequently, potentially disclosing more information about their local data. This raises an underexplored yet critical question: ``\textit{Does improving efficiency through asynchronous aggregation inadvertently increase privacy risk, particularly for high-end clients that participate more often?}''

To rigorously investigate this question under realistic device heterogeneity, we conduct an empirical study using a physical FL testbed spanning five diverse edge devices (Table~\ref{tab:testbed-specs}). This setup captures practical variability in compute capacity and memory—factors that directly affect participation, training dynamics, latency, and privacy exposure. We implement asynchronous FL using staleness-aware \texttt{FedAsync}~\cite{xie2019asynchronous} and compare it to the widely used synchronous baseline, \texttt{FedAvg}.

To formally bound privacy risk from frequent client updates, we incorporate Local Differential Privacy (LDP)\cite{arachchige2019local}, perturbing each client’s update prior to transmission. For precise tracking of cumulative privacy loss during iterative training, we apply the Moments Accountant\cite{abadi2016deep}, enabling fine-grained analysis of how update frequency—particularly under asynchronous coordination—translates into privacy risk over time.

As a benchmark task, we adopt Speech Emotion Recognition (SER), which involves processing speech signals to detect emotional states. SER is both computationally demanding and privacy-sensitive, as it relies on client speech data and is used in applications such as smart homes, automotive systems, and healthcare. This makes it an ideal testbed for evaluating FL under heterogeneous, resource-constrained, and privacy-critical conditions.

Our results reveal substantial trade-offs: asynchronous FL improves efficiency and reduces dependence on slower clients but significantly amplifies privacy risk and fairness disparities. Low-end devices, already limited in compute power, face further marginalization as their updates become stale, exert less influence on the global model, and are more vulnerable to utility loss under LDP noise.

\vspace{0.6em}
Key contributions of our work include:
\begin{itemize}
    \item Empirical comparison of synchronous (\texttt{FedAvg}) and asynchronous (\texttt{FedAsync}) FL under realistic device heterogeneity using a physical edge testbed spanning five hardware tiers.
    \item Fine-grained per-client analysis of cumulative privacy loss with the Moments Accountant, revealing how asynchronous update patterns amplify privacy risks, particularly for high-end clients. 
    \item Evaluation of staleness-aware aggregation in \texttt{FedAsync}, highlighting its overlooked impact on fairness and accuracy under local differential privacy across heterogeneous devices.
    \item Deployment of Speech Emotion Recognition as a computationally demanding, privacy-sensitive benchmark for evaluating FL under heterogeneous, resource-constrained conditions.
    \item Characterization of efficiency–privacy–fairness trade-offs, offering actionable insights for designing efficient and privacy-aware FL systems in heterogeneous environments.
\end{itemize}

\section{Preliminaries and Background}
\label{Sec:prelim}

\subsection{Device Heterogeneity in Federated Learning}
\label{DH-FL}
Device heterogeneity is a fundamental challenge in FL, where participating clients possess diverse computational resources, memory capacities, and network bandwidths. This variation leads to the \textit{straggler effect}, where slower devices delay training in synchronous schemes. Heterogeneity also causes skewed contributions across clients, impacting fairness and convergence rates.

Several strategies have been proposed to address device heterogeneity. Asynchronous FL methods (e.g., \texttt{FedAsync}) allow the server to incorporate client updates as they arrive, mitigating delays from stragglers. These methods often down-weight stale updates to account for delayed contributions. FedBuff~\cite{nguyen2022federated} improves convergence stability by buffering incoming updates before aggregation. Resource-aware client selection schemes, such as FedCS~\cite{nishio2019client} and TiFL~\cite{chai2020tifl}, further enhance efficiency by prioritizing clients based on system capabilities.

However, these methods often prioritize frequent contributions from high-end devices, unintentionally skewing training dynamics and amplifying privacy risks. Moreover, while many prior works attempt to address device heterogeneity in FL, they largely overlook the potential privacy risks arising from heterogeneous client behaviors. Our work explicitly addresses this gap by analyzing privacy trade-offs under synchronous and asynchronous aggregation across real heterogeneous devices.

\subsection{Local Differential Privacy}
\label{Sec:ldp}
Local Differential Privacy (LDP) protects user data by applying random noise directly on the client device before data leaves the device, eliminating the need for a trusted server or aggregator~\cite{arachchige2019local}. In FL, this typically involves perturbing model updates or gradients prior to transmission to the central server.

\textit{Definition 1 (($\epsilon$, $\delta$)-LDP~\cite{bassily2019linear}):} A randomized mechanism $M$ satisfies $(\epsilon, \delta)$-LDP if, for any inputs $v, v'$ and any output $y \in S$,
\[
\Pr[M(v) = y] \leq e^\epsilon \Pr[M(v') = y] + \delta.
\]
Smaller $\epsilon$ implies stronger privacy. In our setting, LDP is enforced using Gaussian noise within the DP-SGD mechanism~\cite{abadi2016deep}.

\subsection{Moments Accountant}
\label{sec:ma}
The Moments Accountant~\cite{abadi2016deep} offers tighter tracking of cumulative privacy loss in iterative training compared to standard composition methods, making it well-suited for FL.

Let $\mathcal{M}$ be a randomized mechanism applied to neighboring datasets $D$ and $D'$. The privacy loss at output $o$ is:
\[
L(o; \mathcal{M}, D, D') = \log \frac{\Pr[\mathcal{M}(D) = o]}{\Pr[\mathcal{M}(D') = o]}.
\]

The $\lambda$-th log moment is defined as:
\[
\mu_{\mathcal{M}}(\lambda) = \log \mathbb{E}_{o \sim \mathcal{M}(D)} \left[ e^{\lambda \cdot L(o)} \right].
\]

If $\mathcal{M}$ is composed of a sequence of mechanisms $\mathcal{M}_1, \ldots, \mathcal{M}_k$, then:
\[
\mu_{\mathcal{M}}(\lambda) \leq \sum_{i=1}^k \mu_{\mathcal{M}_i}(\lambda).
\]

The corresponding $(\epsilon, \delta)$ privacy guarantee is given by:
\[
\delta = \min_{\lambda} \exp\left( \mu_{\mathcal{M}}(\lambda) - \lambda \epsilon \right).
\]

\subsection{Speech Emotion Recognition}
\label{SER-Domain}
SER classifies emotional states from speech signals by analyzing prosody, tone, and spectral features~\cite{george2024review}. It enhances user experiences across domains, including smart homes (Amazon Alexa, Google Nest)~\cite{almusaed2023enhancing}, entertainment\footnote{This is one of the use cases in the DAIS project \url{https://dais-project.eu/}}, healthcare (Fitbit Sense)~\cite{kerkeni2019automatic}, automotive systems (Hyundai's AI-Based Cabin Monitoring System), and education (iPads)~\cite{gupta2023facial}. These applications highlight the versatility of SER across a broad range of hardware, from low-end to high-end devices tailored to domain-specific computational requirements.

Training SER models traditionally requires centralized speech data collection, raising privacy concerns due to the sensitive biometric nature of speech signals~\cite{kroger2020privacy}. FL offers a decentralized alternative, enabling SER model training across distributed heterogeneous devices while preserving data locality~\cite{mohammadi2023balancing}. However, given the heterogeneous nature of client devices across different applications, it is crucial to evaluate potential privacy risks in FL.

% Furthermore, SER models are computationally demanding, requiring substantial CPU and RAM resources, making them well-suited for FL evaluations in heterogeneous environments. This work explores the privacy implications of SER in FL, assessing how heterogeneous client devices and different aggregation methods impact privacy loss.

\section{Methodology}
\label{Sec:methodology}
\begin{figure*}[t]
\centering
\subfloat[Synchronous FL]{%
    \includegraphics[width=3.3in]{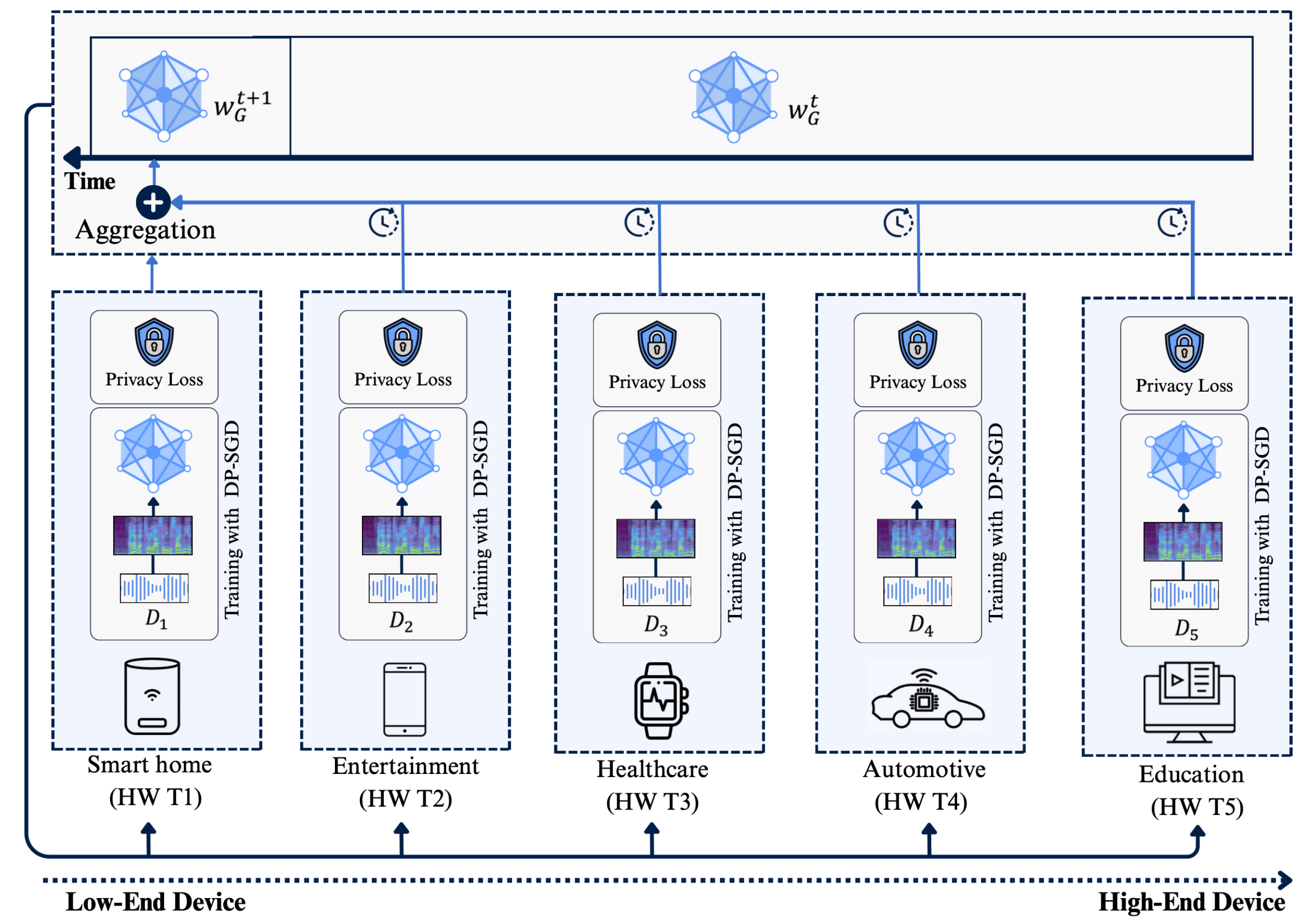}%
    \label{fig:over-asyn}
}
\subfloat[Asynchronous FL]{%
    \includegraphics[width=3.3in]{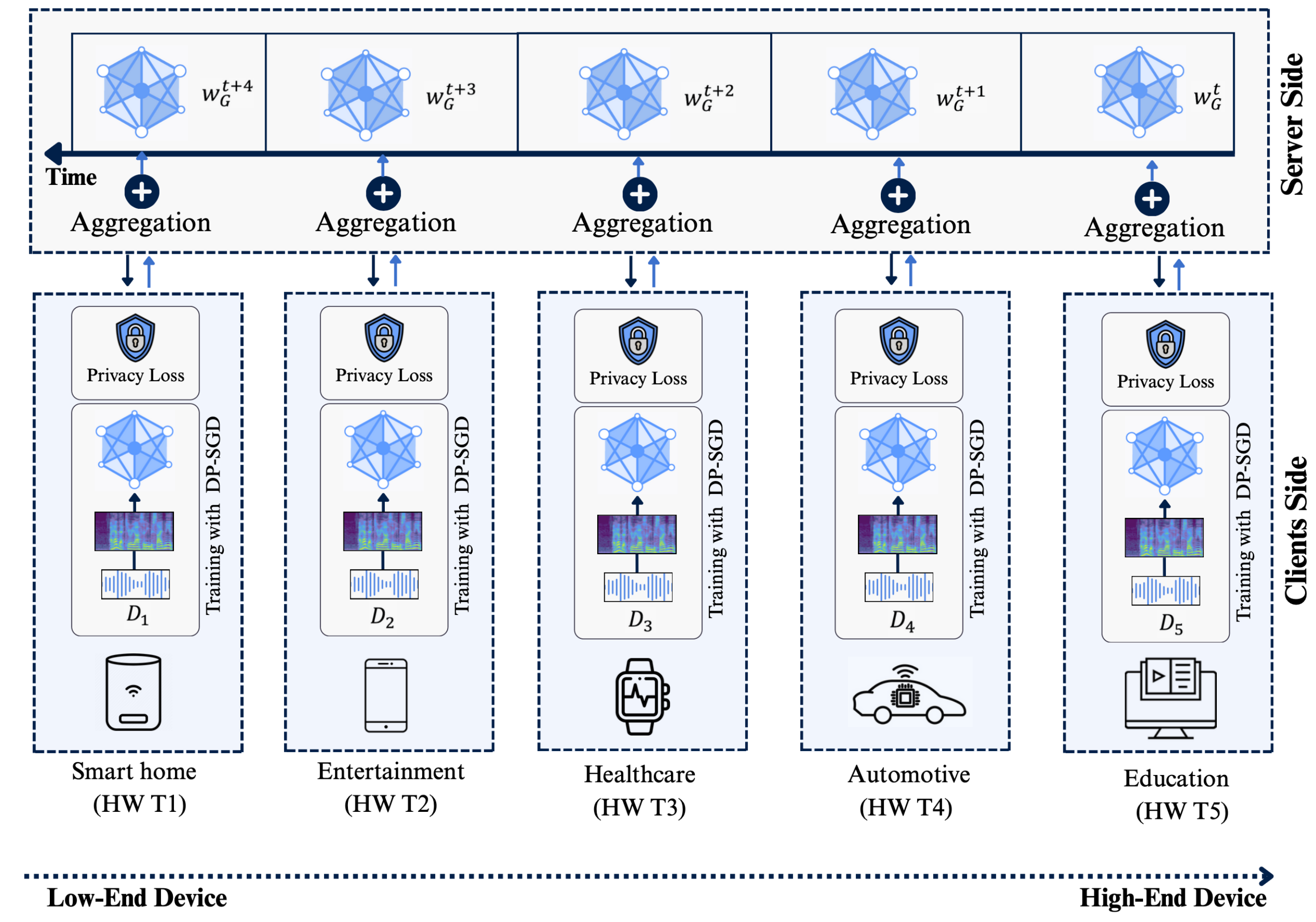}%
    \label{fig:over-sync}
}
\caption{Federated learning under device heterogeneity. Comparison of synchronous and asynchronous FL for speech emotion recognition. Local update privacy is ensured by DP-SGD, with per-client privacy loss tracked by the moments accountant. Devices range from low-end (\textit{HW~$T_1$}) to high-end (\textit{HW~$T_5$}), reflecting realistic hardware variability.}
\label{fig:system}

\end{figure*}

To empirically evaluate the trade-offs between efficiency, fairness, and privacy in synchronous and asynchronous FL under heterogeneous devices for SER, we design a physical FL testbed with five diverse edge devices (Table~\ref{tab:testbed-specs}). These devices are selected to approximate the computational capabilities typically found in real-world SER applications, including smart homes, entertainment, healthcare, automotive, and education. The overall system architecture is shown in Fig.~\ref{fig:system}. Each device maintains a local dataset, denoted as $\mathcal{D}_i$, where $i \in {1, 2, \dots, 5}$.

The global FL objective is to minimize the overall loss across all clients: \begin{equation} w^* = \arg \min_w \sum_{i=1}^5 \frac{|\mathcal{D}_i|}{|\mathcal{D}|} \mathcal{L}_i(w), \end{equation} where $|\mathcal{D}i|$ is the size of client $i$’s dataset, $|\mathcal{D}| = \sum{i=1}^5 |\mathcal{D}_i|$ is the total dataset size, and $\mathcal{L}_i(w)$ is the local loss: \begin{equation} \mathcal{L}_i(w) = \frac{1}{|\mathcal{D}i|} \sum{j=1}^{|\mathcal{D}_i|} \ell(f_w(x_j), y_j), \label{local-eq} \end{equation} with $f_w(x_j)$ as the model prediction, $y_j$ as the ground truth, and $\ell(\cdot, \cdot)$ denoting the cross-entropy loss.

Our framework implements both synchronous (\texttt{FedAvg}) and asynchronous (\texttt{FedAsync}) aggregation. To ensure local privacy, each client applies DP-SGD~\cite{abadi2016deep}, and cumulative privacy loss $\epsilon$ is tracked using the Moments Accountant~\cite{abadi2016deep, geyer2017differentially}.

\subsection{Speech Emotion Recognition Task}
\label{usecase}
SER aims to classify emotional states from spoken utterances by analyzing both temporal and spectral patterns in the audio signal. We use mel-spectrograms as input features and a lightweight Convolutional Neural Network (CNN) for classification.\\

\noindent \textbf{Feature Extraction.} Each client converts raw audio into mel-spectrograms \( S_{\text{mel}}(t, f) \)~\cite{issa2020speech}, capturing time-frequency characteristics. This is computed via the Short-Time Fourier Transform (STFT) followed by a mel filter bank:

\begin{equation}
    S_{\text{mel}}(t, f) = \sum_{k=1}^{N} |X(t, f_k)|^2 H_{\text{mel}}(f_k),
    \label{eq:mel_spectrogram}
\end{equation}
where \( X(t, f_k) \) is the STFT of the input signal \( x(t) \), and \( H_{\text{mel}}(f_k) \) is the mel-scale filter response.\\

\noindent \textbf{Model Architecture.} We adopt a compact CNN tailored for SER on resource-constrained devices, inspired by prior work~\cite{aftab2021lightsernet, simic2024enhancing}. The architecture processes 1D feature maps derived from mel-spectrograms and consists of:

\begin{itemize}
    \item Two 1D convolutional layers with 64 and 128 filters (kernel size 5), each followed by Group Normalization and ReLU activation.
    \item 1D max-pooling layers (pool size 2) after each convolution block to reduce feature map size.
    \item Dropout layers after convolution (rates 0.3 and 0.4) and fully connected layers (rate 0.5) to enhance generalization.
    \item A fully connected layer with 128 units, followed by an output layer for emotion classification.
\end{itemize}
The model is trained using the Adam optimizer (learning rate 0.001) and cross-entropy loss (Eq.~\eqref{local-eq}). Its lightweight design offers a balance of accuracy and efficiency, making it well-suited for FL deployment across heterogeneous devices.

\subsection{Federated Learning Setup}
\label{FL}
The following sections, along with Algorithm~\ref{alg:fl_system_ser}, summarize the FL workflow, detailing both client-side and server-side operations.

\paragraph{Client-Side}
\hspace{1em}

\noindent \textbf{Local Training.} Each client \( C_k \) receives the current global model \( W_G^t \) and trains locally on its dataset \( D_k \), preprocessed into mel-spectrograms \( S_{\text{mel}}(t, f) \). Training is performed for \( E \) local epochs using mini-batches of size \( B \) and follows the DP-SGD procedure. \\

\noindent \textbf{Local Differential Privacy.} For each mini-batch, per-sample gradients are clipped and noised to enforce LDP:
\begin{equation}
g_i \leftarrow \frac{g_i}{\max\left(1, \frac{\|g_i\|_2}{C}\right)},
\end{equation}
then noise is added:
\begin{equation}
\tilde{g} = \frac{1}{|b|} \sum_{i \in b} g_i + \mathcal{N}(0, \sigma^2 C^2 I),
\end{equation}
and the local model is updated:
\begin{equation}
W_k \leftarrow W_k - \eta \tilde{g}.
\end{equation}

% \noindent \textbf{Local Differential Privacy.} For each mini-batch, per-sample gradients are clipped and noised to enforce LDP:
% \begin{equation}
% g_i \leftarrow g_i \cdot \min\left(1, \frac{S}{\|g_i\|_2}\right),
% \end{equation}
% noise is added:
% \begin{equation}
% \tilde{g} = \frac{1}{|b|} \sum_{i \in b} g_i + \mathcal{N}(0, \sigma^2 S^2 I),
% \end{equation}
% then the local model is updated with:
% \begin{equation}
% W_k \leftarrow W_k - \eta \tilde{g}.
% \end{equation}

% The updated model \( W_k \) is then sent to the server for aggregation.\\

\noindent \textbf{Privacy Loss Accounting.} 
We adopt the Moments Accountant framework~\cite{abadi2016deep} to track cumulative privacy loss under device heterogeneity. This method provides tighter bounds than standard composition and is well-suited to settings with non-uniform client participation, such as \texttt{FedAsync}~\cite{hu2020personalized}.

Each client tracks its privacy loss locally. We fix the noise scale \( \sigma \), clipping norm \( C \), and define the sampling probability \( q = \frac{B}{|\mathcal{D}_k|} \), then compute the cumulative privacy budget \( \epsilon_k^t \) over training rounds.

At each round \( t \), the privacy log moment is:
\begin{equation}
\mu_t(\lambda) = \log \mathbb{E} \left[ \left( \frac{\mathcal{M}(D)}{\mathcal{M}(D')} \right)^\lambda \right],
\end{equation}
where \( \mathcal{M} \) is the Gaussian mechanism, and \( D \), \( D' \) are neighboring datasets differing by one client’s data.

The cumulative moment and resulting privacy budget are:
\begin{equation}
\mu(\lambda) = \sum_{t=1}^{T} \mu_t(\lambda), \quad
\epsilon = \min_{\lambda} \left( \frac{\mu(\lambda) - \log \delta}{\lambda} \right).
\end{equation}

This formulation enables per-client, round-level privacy tracking and supports fair comparison across aggregation strategies.

\begin{algorithm}[t]
\caption{FL for SER with \texttt{FedAvg} / \texttt{FedAsync}, incorporating LDP and privacy accounting}
\label{alg:fl_system_ser}
\small
\begin{algorithmic}[1]
\Require Global model $W_G^0$, total rounds $T$, clients $K$, learning rate $\eta$, clipping norm $C$, noise scale $\sigma$, failure probability $\delta$, base decay factor $\alpha$, batch size $B$, local epochs $E$

\While{$t \leq T$ \textbf{or until convergence}}
    \State Select clients $\mathcal{S}^t = \{C_1, \dots, C_K\}$
    \For{each client $C_k \in \mathcal{S}^t$ \textbf{in parallel}}
        \State \textbf{Client-side:}
        \State Receive $W_G^t$ and set $W_k \gets W_G^t$
        \For{epoch $e = 1,\dots,E$}
            \For{mini-batch $b \in B$}
                \State Compute per-sample gradients $\{g_i\}_{i \in b}$
                \State Clip: $g_i \leftarrow \frac{g_i}{\max\left(1, \frac{\|g_i\|_2}{C}\right)}$
                \State Perturb: $\tilde{g} \leftarrow \frac{1}{|b|} \sum_{i \in b} g_i + \mathcal{N}(0, \sigma^2 C^2 I)$
                \State Update: $W_k \leftarrow W_k - \eta \tilde{g}$
            \EndFor
        \EndFor

        \State \textbf{Privacy Accountant:}
        \State Compute $\mu_t^{(k)}(\lambda)$ 
        \State Update $\mu^{(k)}(\lambda) \gets \mu^{(k)}(\lambda) + \mu_t^{(k)}(\lambda)$
        \State $\epsilon_k^t \gets \min_\lambda \left( \frac{\mu^{(k)}(\lambda) - \log \delta}{\lambda} \right)$

        \State Send $W_k$ and timestamp $t_k$ to server
    \EndFor

    \State \textbf{Server-side:}
    \If{\texttt{FedAvg}}
        \State $p_k \gets \frac{N_k}{\sum_{j \in \mathcal{S}^t} N_j}$
        \State $W_G^{t+1} \gets \sum_{k \in \mathcal{S}^t} p_k W_k$
    \Else \Comment{FedAsync}
        \For{each received update $(W_k, t_k)$}
            \State Compute staleness $\tau_k = t - t_k$
            \State $\alpha_k \gets \frac{\alpha}{1 + \tau_k}$
            \State $W_G^{t+1} \gets (1 - \alpha_k) W_G^t + \alpha_k W_k$
        \EndFor
    \EndIf
\EndWhile

\State \Return Final global model $W_G^T$, per-client privacy budgets $\{\epsilon_k^T\}$
\end{algorithmic}
\end{algorithm}

\paragraph{Server-Side}
The server aggregates received updates using one of the following strategies:

\noindent \textbf{Synchronous (\texttt{FedAvg}).} Updates are aggregated only after all selected clients return their models as follows: 
\begin{equation}
    W_G^{t+1} = \sum_{k \in \mathcal{S}^t} p_k W_k, \quad 
    \label{eq:fedavg_update_ser}
\end{equation}
where \( p_k = \frac{N_k}{\sum_{j \in \mathcal{S}^t} N_j} \) is the client weighting based on local dataset size.

\noindent \textbf{Asynchronous (\texttt{FedAsync}).} As updates arrive from clients, the server applies them immediately. To account for delayed updates (staleness), we adopt a staleness-aware decay factor~\cite{xie2019asynchronous}:
\begin{equation}
    \alpha_k = \frac{\alpha}{1 + \tau_k},
    \label{eq:fedasync_alpha}
\end{equation}
\begin{equation}
    W_G^{t+1} = (1 - \alpha_k) W_G^t + \alpha_k W_k,
    \label{eq:fedasync_update_ser}
\end{equation}
where $\tau_k$ denotes the staleness of client $C_k$'s update, and $\alpha \in (0, 1]$ is a decay factor that down-weights older updates to stabilize training. Larger $\alpha_k$ favors fresher updates for faster adaptation, while smaller values promote conservative, stable convergence under heterogeneity.

\section{Experimental Design and Evaluation Result}
This section outlines the experimental design (Section~\ref{sec:ED}) and evaluation result (Section~\ref{ER}).

\subsection{Experimental Design}
\label{sec:ED}
 This section presents the experimental setup for our study.

\subsubsection{Hardware Configuration of FL Testbed}\label{sec:device}
\begin{wrapfigure}{r}{0.42\textwidth}
    \vspace{-20pt}
    \centering
    \includegraphics[width=0.3\textwidth]{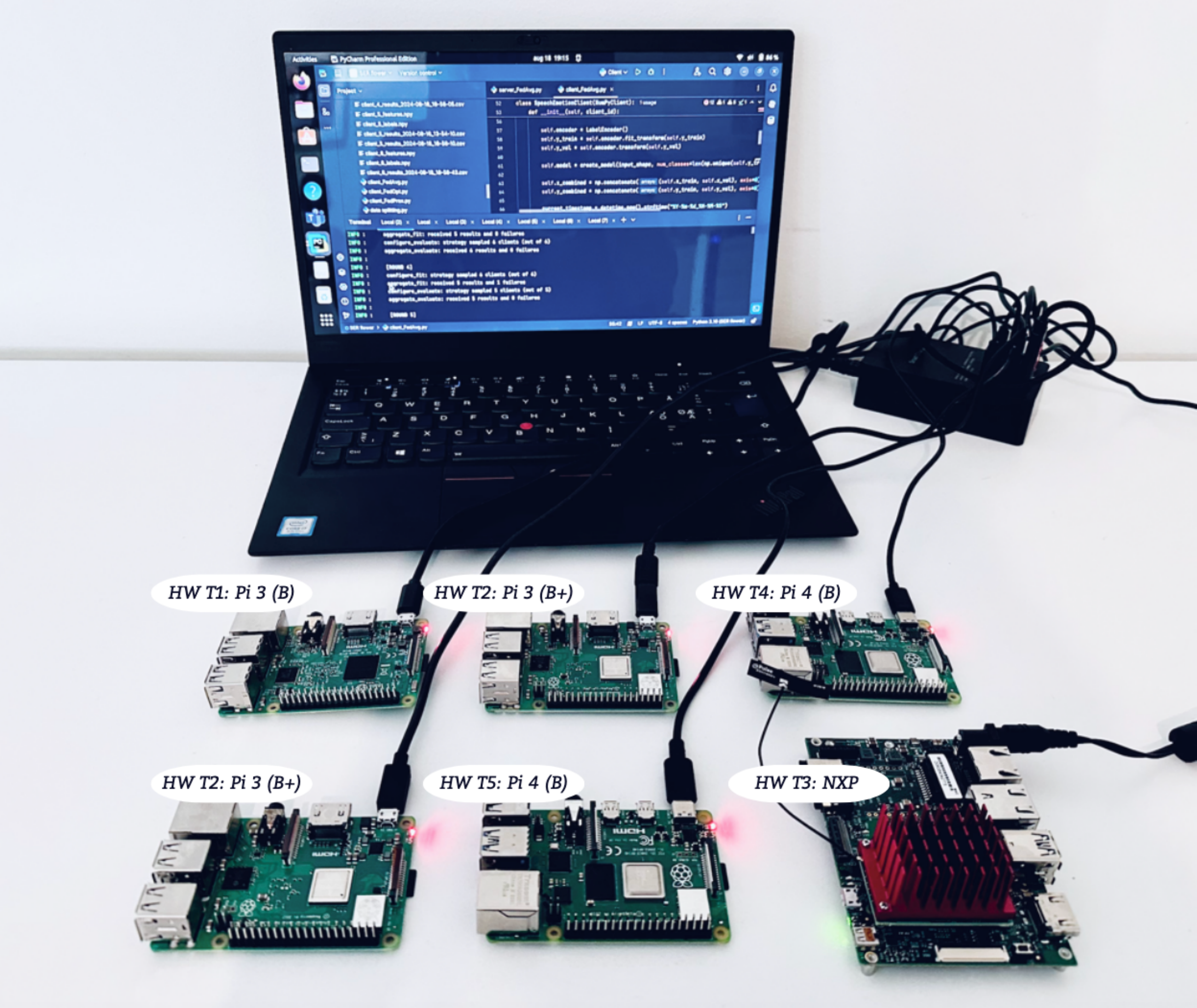}
    \caption{FL testbed with clients comprising heterogeneous devices (HW~$T_1$--$T_5$)}
    \label{fig:device}
    \vspace{-10pt}
\end{wrapfigure}
To accurately reflect real-world SER deployment scenarios (as outlined in Section~\ref{SER-Domain}), we built an FL testbed comprising five heterogeneous edge devices and a central server. The devices span five distinct hardware configurations (HW~$T_1$--$T_5$), varying in computational capacity (CPU, RAM), as detailed in Table~\ref{tab:testbed-specs} and shown in Fig.\ref{fig:device}.\footnote{Two devices shared the HW$T_2$ configuration, but only one was used as a unique client in the experiments.}

The edge devices include Raspberry Pi 3 Model B and B+, Raspberry Pi 4 Model B (4GB, 8GB), and an NXP HummingBoard, each equipped with a 32GB SD card for local data storage. The central server is a ThinkPad X1 Carbon with an Intel i7 CPU and 16GB RAM. All devices communicate over a 2.4GHz Wi-Fi network, emulating realistic wireless conditions in edge computing environments.

\begin{table*}[!t]
    \centering
    \captionsetup{justification=centering, font=small, labelfont=bf, labelsep=period, position=above}
    \caption{Heterogeneous edge device specifications and closely mapped hardware types for SER Applications}
    \label{tab:testbed-specs}
    \renewcommand{\arraystretch}{1.2} % Increase row height by 1.3 times
    \begin{threeparttable}
    \scalebox{0.7}{
        \begin{tabular}{| p{1.1cm}| p{3.5cm}| p{6cm} | c | p{4.2cm}| p{3.8cm}|}
            \hline
            \textbf{Hardware type} & \textbf{Device} & \textbf{CPU} & \textbf{RAM} & \textbf{Wireless specification} & \textbf{Closely Map to associated SER application} \\
            \hline
            HW $T_1$ & Raspberry Pi 3 Model B & 1.2GHz quad-core ARM Cortex-A53 & 1GB & 2.4GHz Wi-Fi & Smart Homes (Lowe-end) \\
            \hline
            HW $T_2$ & Raspberry Pi 3 Model B+ & 1.4GHz quad-core ARM Cortex-A53 & 1GB & Dual-band 2.4GHz and 5GHz Wi-Fi & Entertainment (Low-Mid Range) \\
            \hline
            HW $T_3$ & NXP device\tnote{*} & 1.5GHz-1.8GHz dual to quad-core ARM Cortex-A53 & 1GB & Dual-band 2.4GHz and 5GHz Wi-Fi & Healthcare (Moderate) \\
            \hline
            HW $T_4$ & Raspberry Pi 4 Model B (4GB) & 1.5GHz quad-core ARM Cortex-A72 & 4GB & Dual-band 2.4GHz and 5GHz Wi-Fi & Automotive (High-Mid Range) \\
            \hline
            HW $T_5$ & Raspberry Pi 4 Model B (8GB) & 1.5GHz quad-core ARM Cortex-A72 & 8GB & Dual-band 2.4GHz and 5GHz Wi-Fi & Education (High-End) \\
            \hline
        \end{tabular}
    }
    \begin{tablenotes}
        \footnotesize
        \item[*] NXP HummingBoard provided through DAIS project partnership. More at: \url{https://developer.solid-run.com}
    \end{tablenotes}
    \end{threeparttable}
     \vspace{-2pt}
\end{table*}

% \begin{figure}[!t]
%     \centering
%     \captionsetup{justification=centering}
%     \includegraphics[width=6cm]{Figure/testbed1.pdf}
%     \caption{FL testbed with clients comprising heterogeneous devices (HW~$T_1$--$T_5$)}
%     \label{fig:device}
%     \vspace{-4pt}
% \end{figure}

\subsubsection{Software Environment}
\label{Soft-Env}
Client devices ran Ubuntu 22.04.4 LTS (Raspberry Pis) or Debian (NXP HummingBoard). The FL framework was implemented using Flower~\cite{beutel2020flower}, with LDP enforced via Opacus~\cite{yousefpour2021opacus} integrated into PyTorch. As Opacus does not natively support the Moments Accountant, we developed a custom module to track cumulative privacy loss across rounds.

\subsubsection{Dataset}
We used the CREMA-D dataset~\cite{cao2014crema}, a widely adopted benchmark for speech emotion recognition (SER), comprising 7,442 audio clips from 91 actors across six emotions. For this study, we focused on the four most frequent classes—\textit{Neutral}, \textit{Happy}, \textit{Angry}, and \textit{Sad}—yielding 5,882 samples after excluding \textit{Fear} and \textit{Disgust}. The data was split into five IID partitions (one per client), each with an 80/20 train-test split, ensuring balanced class distributions for isolating device heterogeneity effects.

\subsubsection{Parameter Settings}
\label{param-settings}
We used \(K = 5\) clients mapped to distinct hardware configurations (Table~\ref{tab:testbed-specs}). Each client received approximately 941 training and 234 test samples. Local training used batch size \(B = 128\), one epoch per round (\(E = 1\)), and learning rate \(\eta = 0.001\). Global training proceeded until convergence (about 60 rounds for \texttt{FedAvg}; continuous updates for \texttt{FedAsync}). We set the clipping (\(C = 1\)), failure probability \(\delta = 10^{-5}\), and Gaussian noise \(\sigma \in \{0.5, 1, 1.5, 2\}\). The sampling probability \(q = \frac{B}{|\mathcal{D}_k|} \approx 0.136\) was used in the moments accountant. For \texttt{FedAsync}, we evaluated decay factors \(\alpha \in \{0.2, 0.4, 0.6\}\), with update weights \(\alpha_k = \frac{\alpha}{1 + \tau_k}\), where \(\tau_k = t - t_k\) denotes staleness. Results were averaged over 10 random seeds.

\subsection{Evaluation Results}
\label{ER}
Section~\ref{eval:eff-per} analyzes per-client performance and efficiency under different aggregation modes and device heterogeneity, followed by Section~\ref{eval:fair-tradeoff}, which examines fairness under varying $\alpha$ values. Section~\ref{sec:privacy-analysis} presents per-client privacy loss and corresponding accuracy degradation due to LDP. Finally, Section~\ref{eval:trade-off} discusses the overall trade-offs across these dimensions in heterogeneous FL.\\

\subsubsection{Performance and Efficiency Analysis}\label{eval:eff-per}

~

\noindent \textbf{Resource efficiency across devices.}   
We evaluate how device-level computational and memory constraints affect participation reliability. Table~\ref{tab:resource} reports average RAM usage and cumulative CPU time over 60 training rounds, using \texttt{FedAvg} to ensure consistent participation across devices. Notably, low-end devices occasionally dropped out and rejoined during training.

\begin{wraptable}{r}{0.55\textwidth}
\vspace{-10pt}
\centering
\caption{Approximate resource utilization for each hardware type during FL training (\(T = 60\)), including CPU time (user/system) and RAM usage.}
\label{tab:resource}
\renewcommand{\arraystretch}{1.3}
\scalebox{.8}{
\begin{tabular}{c c c c}
\hline
\multirow{2}{*}{\textbf{HW Type}} & \multicolumn{2}{c}{\textbf{CPU Time (s)}} & \multirow{2}{*}{\textbf{RAM Usage (\%)}}\\
 \cmidrule(lr){2-3} 
 & \textbf{\textit{User}} & \textbf{\textit{System}} &  \\
\hline
\textit{HW~$T_1$} & \(\mathbf{2268.2 \pm 95.4}\) & \(\mathbf{311.0 \pm 28.7}\) & \( \mathbf{78.7 \pm 2.1}\) \\
\hline
\textit{HW~$T_2$} & \(2087.9 \pm 84.6\) & \(275.2 \pm 24.5\) & \(77.1 \pm 1.8\) \\
\hline
\textit{HW~$T_3$} & \(1117.3 \pm 32.1\) & \(93.7 \pm 7.2\) & \(77.0 \pm 1.2\) \\
\hline
\textit{HW~$T_4$} & \(1122.0 \pm 21.4\) & \(83.3 \pm 6.1\) & \(49.6 \pm 1.0\) \\
\hline
\textit{HW~$T_5$} & \(\mathbf{1036.4 \pm 18.9}\) & \(\mathbf{80.9 \pm 5.3}\) & \(\mathbf{30.5 \pm 0.7}\) \\
\hline
\end{tabular}}
% \vspace{-1pt}
\end{wraptable}

\textit{HW~$T_1$} (1GB RAM) reached 78.7\% RAM usage and 2579~s CPU time, with three observed dropouts; \textit{HW~$T_2$} showed similar behavior with two dropouts. Both frequently acted as \textit{stragglers}, delaying aggregation and slowing convergence, which limits their suitability for smart home and entertainment applications.

\textit{HW~$T_3$}, with comparable RAM but improved CPU, maintained stable participation without dropouts, making it well suited for healthcare use. High-end devices (\textit{HW~$T_4$}, \textit{HW~$T_5$}) exhibited low resource usage (approximately 49.6\% and 30.5\% RAM) and minimal CPU time, with no dropouts observed. Their efficiency and stability, make them strong candidates for real-time SER in automotive and educational domains.\\

\noindent\textbf{FL training performance across devices.}\label{eval:performance}

\begin{figure*}[t!]
\centering
\subfloat[Test accuracy]{\includegraphics[width=2.2in]{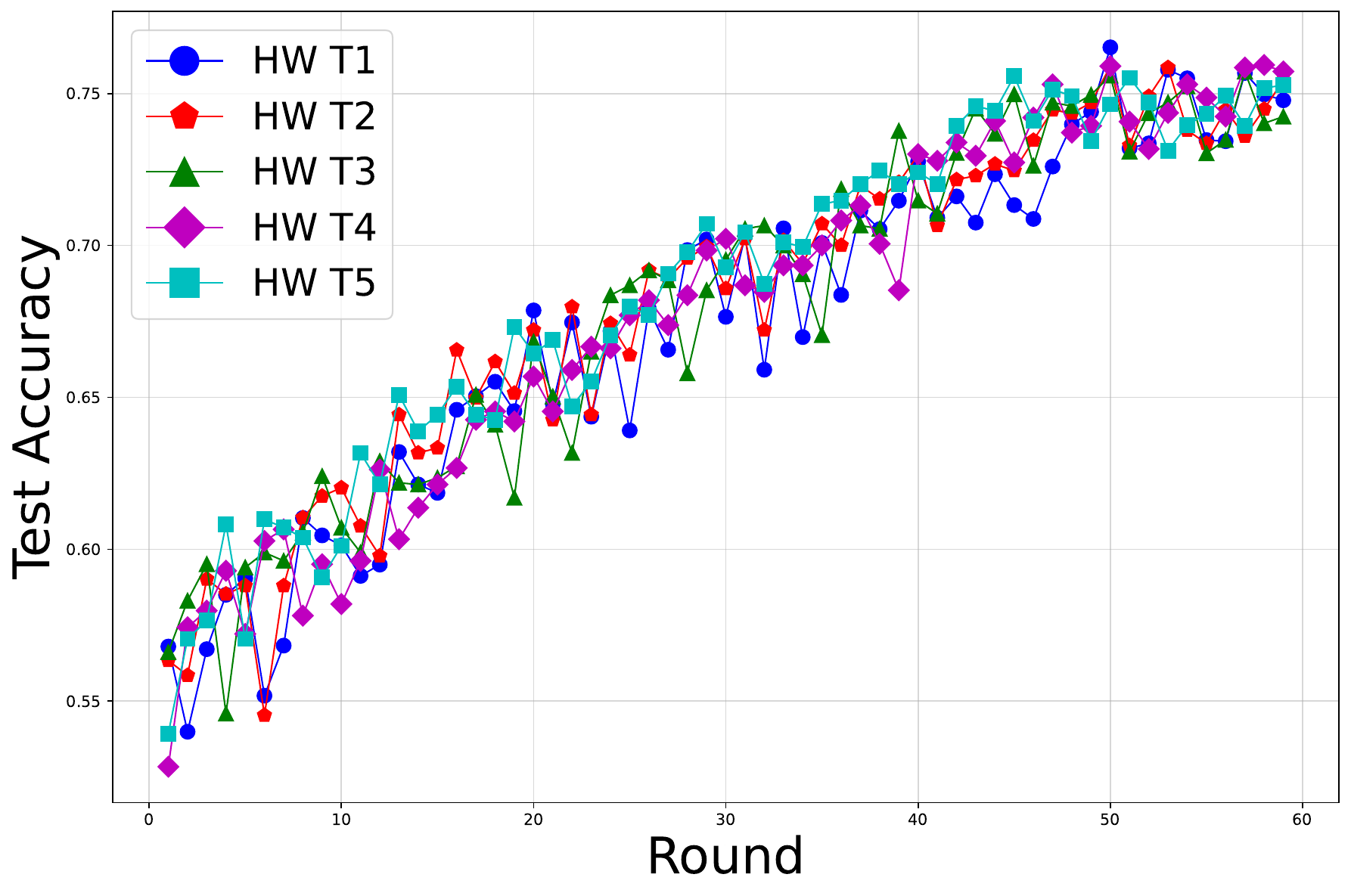}%
\label{fig:acc}}
\subfloat[Training time]{\includegraphics[width=2.2in]{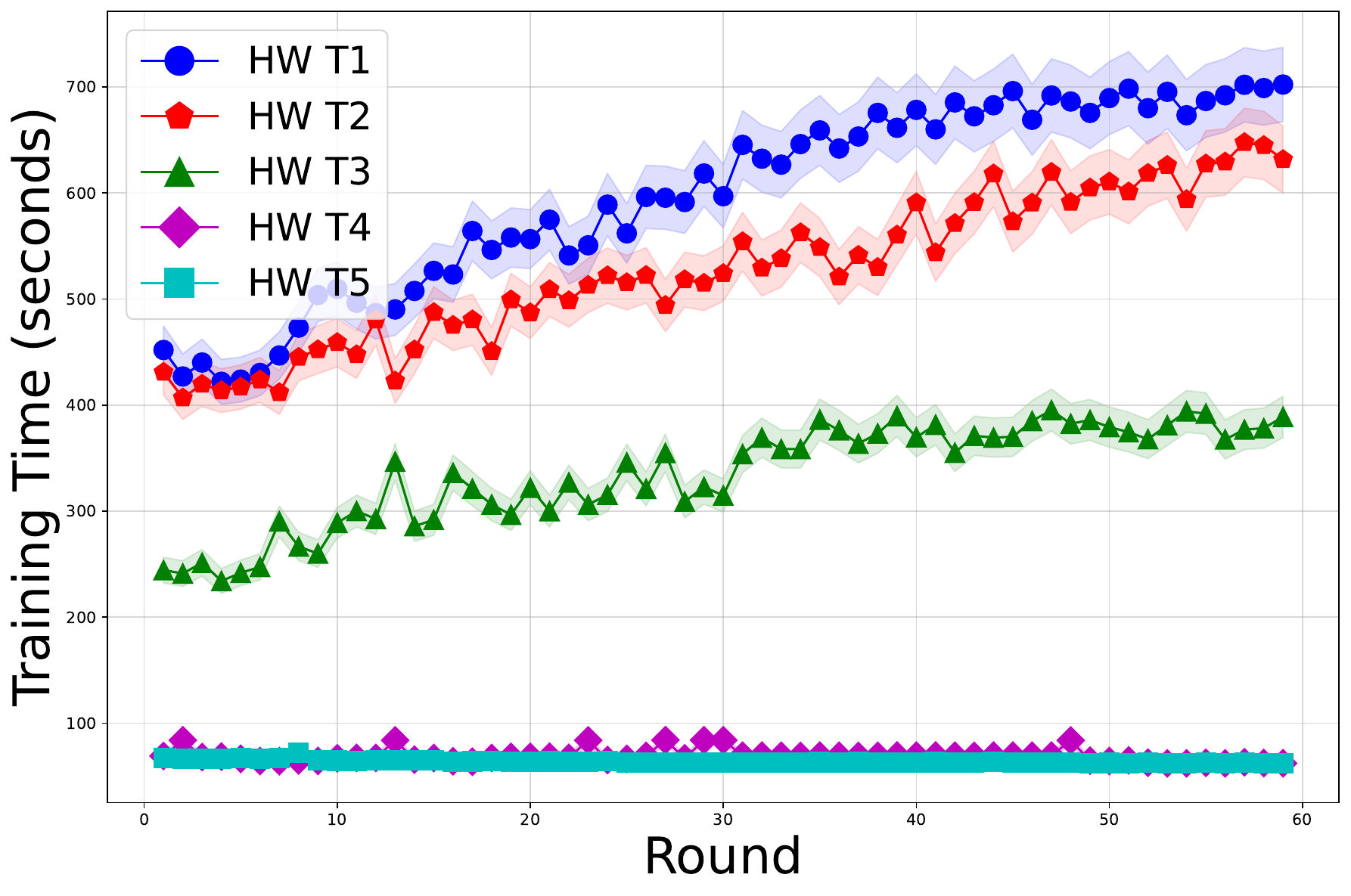}%
\label{fig:trainingtime}}
\subfloat[Update exchange time]{\includegraphics[width=2.2in]{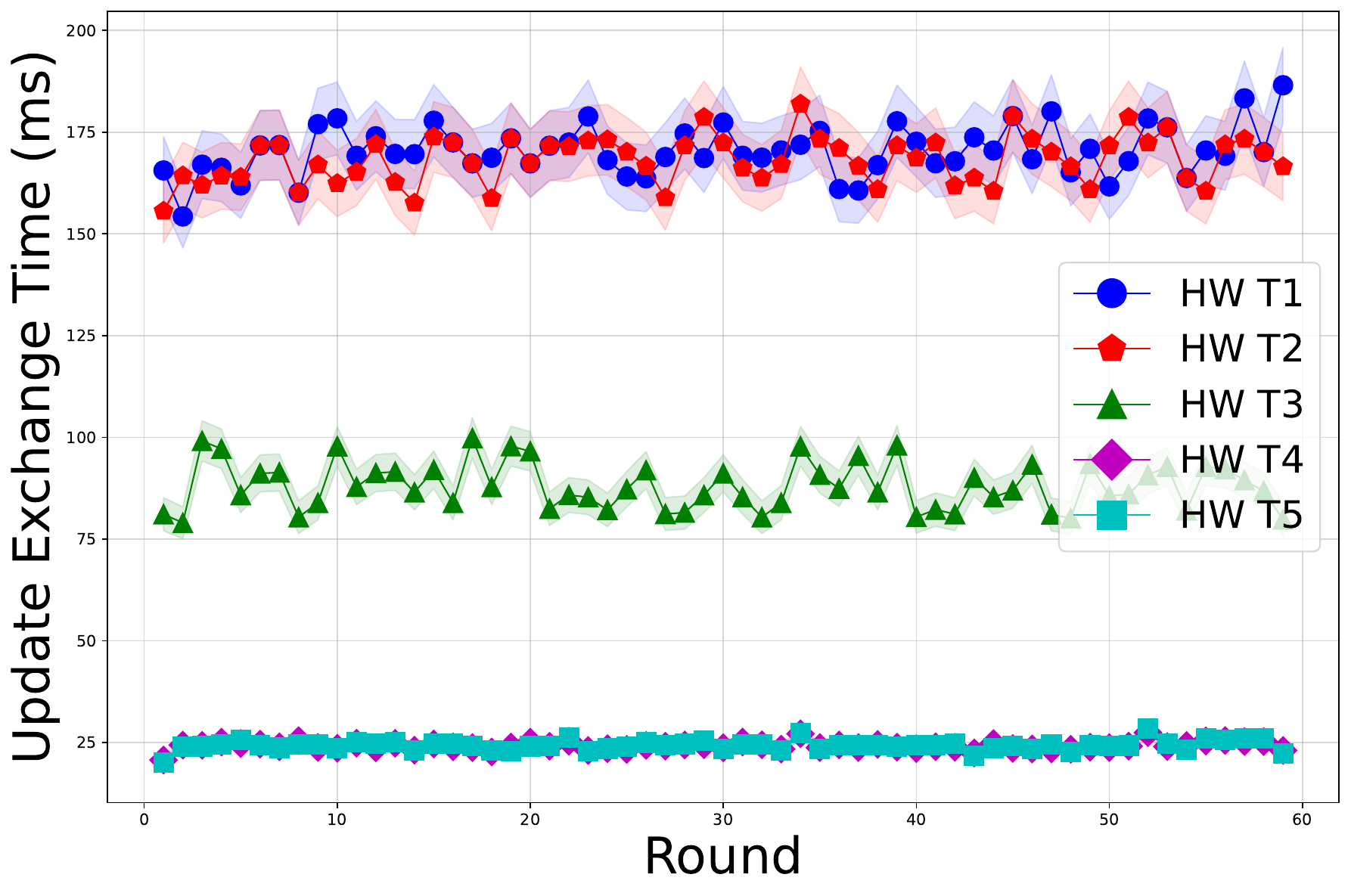}%
\label{fig:latency}}
\caption{FL training performance per round across heterogeneous devices.}
\label{fig:overall-perf}
\vspace{-5pt}
\end{figure*}

We then evaluate training performance by measuring test accuracy (Fig.~\ref{fig:acc}), training time (Fig.~\ref{fig:trainingtime}), and update exchange latency (Fig.~\ref{fig:latency}) across devices. Low-end devices (\textit{HW~$T_1$}, \textit{HW~$T_2$}) showed 3–4$\times$ higher accuracy variance and 6–9$\times$ longer training times, largely due to resource constraints and periodic dropouts and rejoining. Their exchange latency was about 7$\times$ higher, increasing server wait times and slowing convergence. While the model reached 75\% test accuracy after 60 rounds, the intermittent participation of low-end clients reduced their influence on the global model. Notably, despite IID splits, SER remains challenging due to speaker- and emotion-specific variability, making performance particularly sensitive to participation gaps.

\textit{HW~$T_3$} demonstrated intermediate performance, with training time and exchange latency approximately 3–4$\times$ faster than low-end devices and 3–4$\times$ slower than high-end devices. High-end devices (\textit{HW~$T_4$}, \textit{HW~$T_5$}) achieved the best results, showing smooth accuracy improvements with minimal test accuracy fluctuations, training durations of 65–75~s, and exchange latencies near 25~ms, meeting real-time requirements for complex SER tasks.\\

\begin{wrapfigure}{r}{0.52\textwidth}
    \vspace{-5pt}  % Optional: moves figure up
    \centering
    \includegraphics[width=0.46\textwidth]{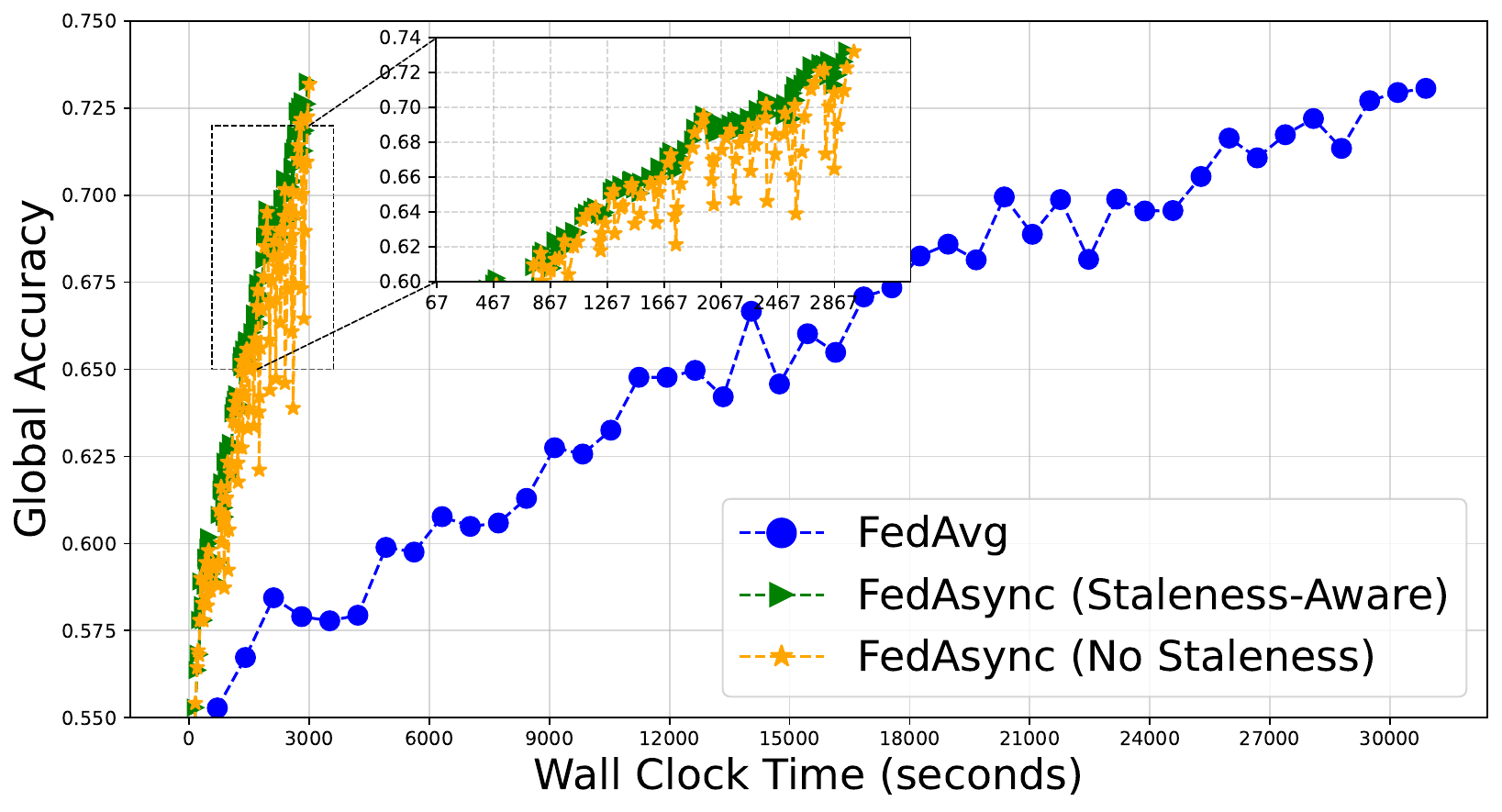}
    \caption{Convergence time of \texttt{FedAvg} vs. \texttt{FedAsync} with and without staleness-aware aggregation under device heterogeneity.}
    \label{fig:async-sync}
    \vspace{-4pt}  % Optional: adjust space below
\end{wrapfigure}

\noindent\textbf{Convergence time by aggregation mode.}\label{eval:wall-time}
We also evaluate the impact of aggregation mode on convergence time under device heterogeneity. As shown in Fig.~\ref{fig:async-sync}, \texttt{FedAvg} requires approximately 27{,}000~s to reach 75\% accuracy. In contrast, \texttt{FedAsync} converges $\sim$9--10$\times$ faster, reaching the same accuracy within $\sim$2{,}900~s at $\alpha = 0.4$, by mitigating the \textit{stragglers effect} and allowing fast devices to advance training without waiting for slower clients.

Notably, we also observe the role of staleness-aware aggregation. While both \texttt{FedAsync} variants achieve substantial speedups, staleness-aware \texttt{FedAsync} delivers smoother convergence with fewer fluctuations. Without staleness control, asynchronous updates exhibit larger instabilities due to the destabilizing influence of outdated updates, particularly from slower devices. Empirically, average round gaps (\textit{i.e.}, staleness) are $\tau \approx 0$ for \textit{HW~$T_5$} and \textit{HW~$T_4$}, but increase to $\tau = 4, 6, 7$ for \textit{HW~$T_3$}, \textit{HW~$T_2$}, and \textit{HW~$T_1$}, respectively. By scaling contributions as $\alpha_k = \frac{\alpha}{1 + \tau}$, staleness-aware aggregation effectively reduces the impact of outdated updates, improving stability without sacrificing convergence speed.\\

\subsubsection{Fairness in Client Participation}
\label{eval:fair-tradeoff}
\begin{figure*}[t]
\centering
\subfloat[$\alpha = 0.2$, staleness-aware]{\includegraphics[width=2.2in]{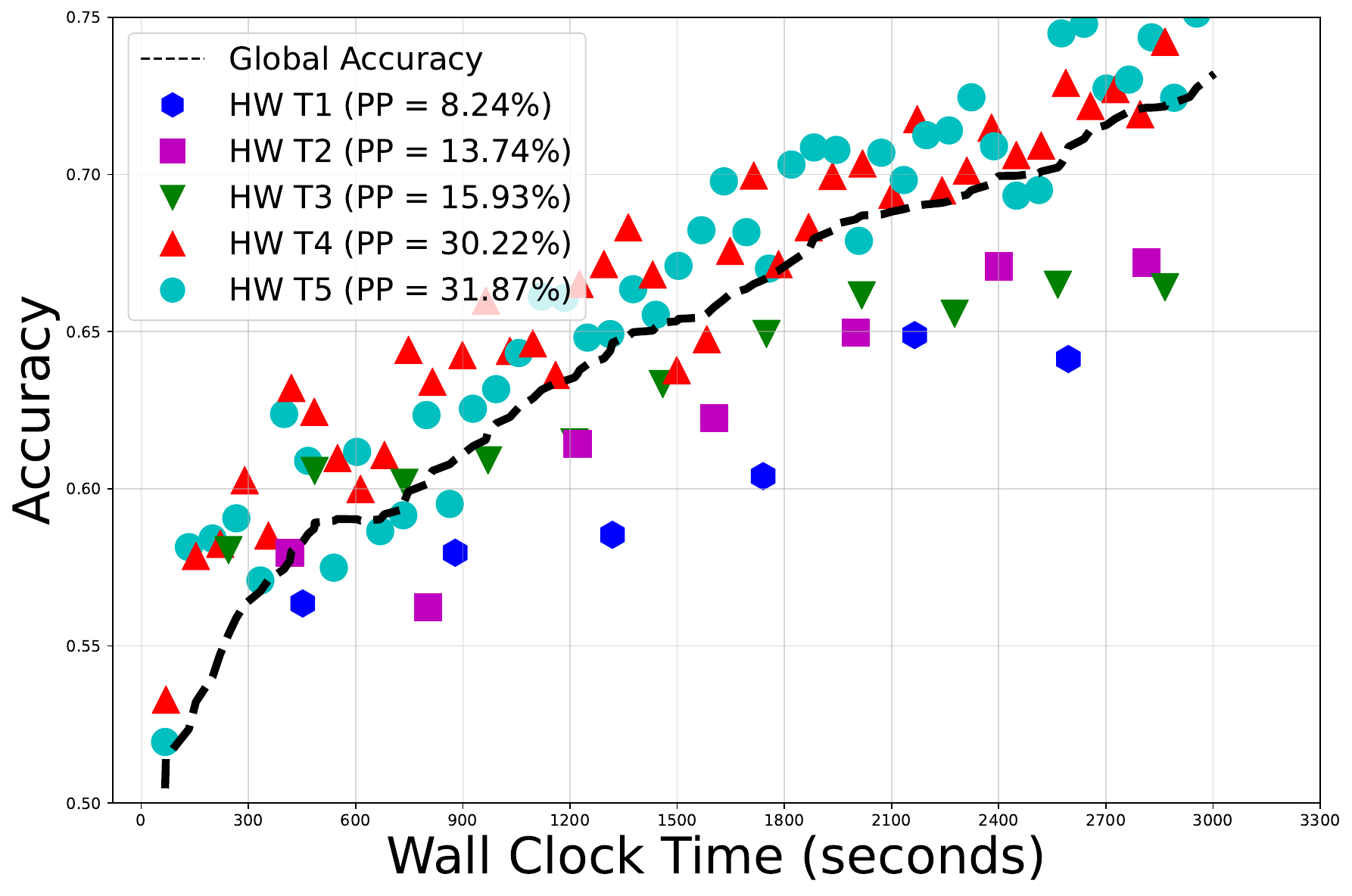}%
\label{fig:async2}}
\subfloat[$\alpha = 0.4$, staleness-aware]{\includegraphics[width=2.2in]{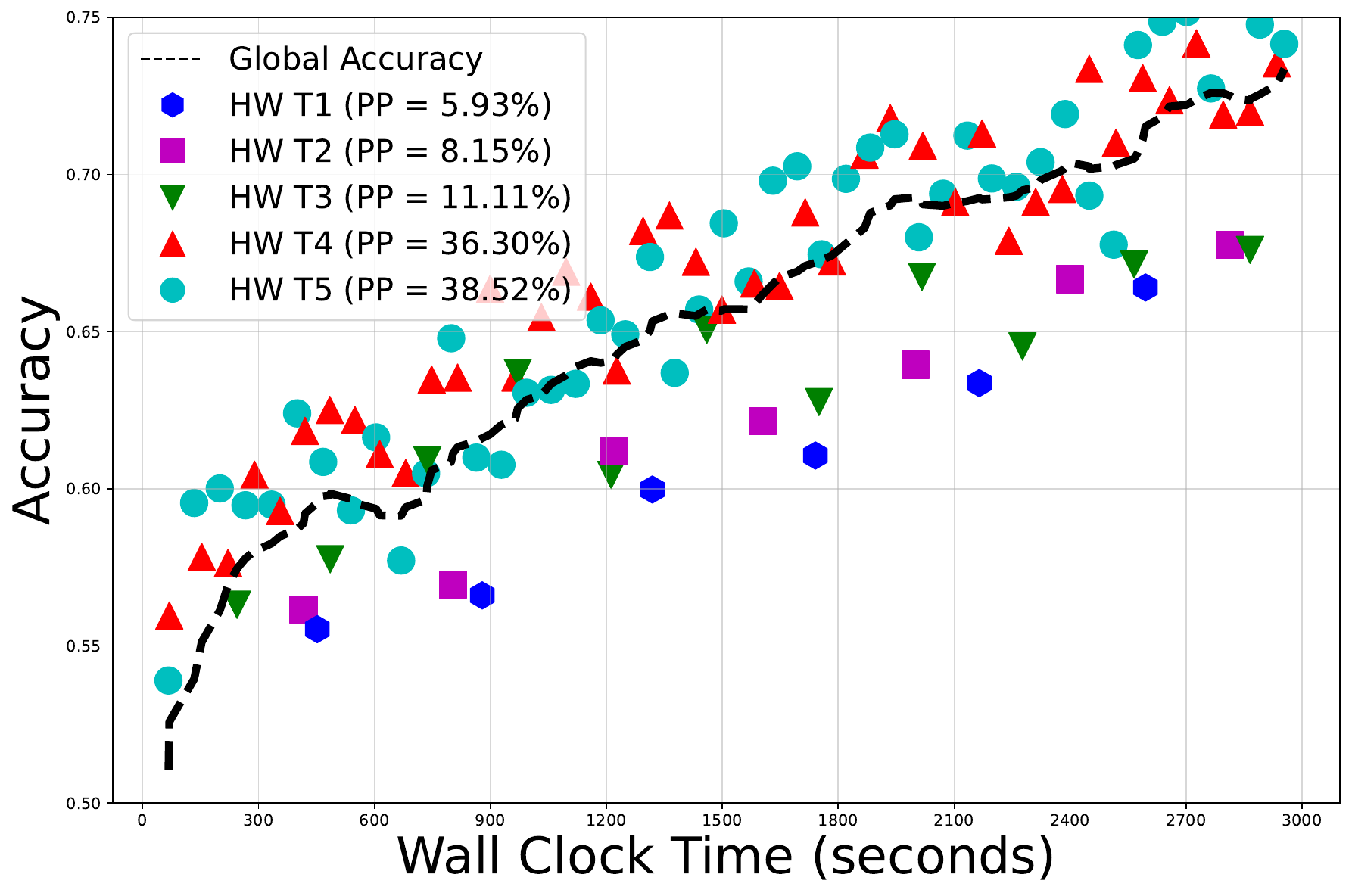}%
\label{fig:async4}}
\subfloat[$\alpha = 0.6$, staleness-aware]{\includegraphics[width=2.2in]{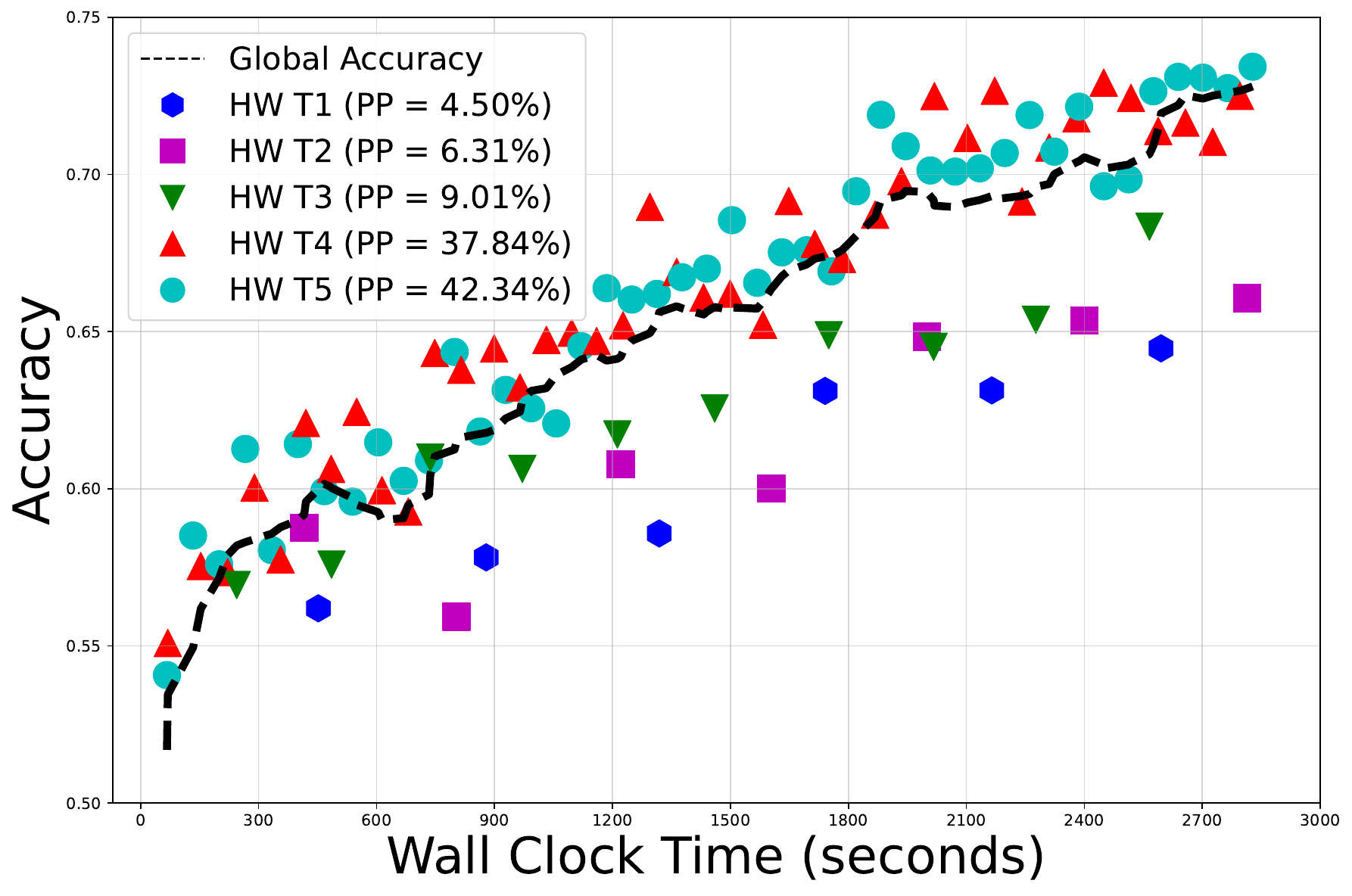}%
\label{fig:async6}}
\caption{Global and per-device accuracy trajectories, along with participation percentages (PP), in asynchronous FL with staleness-aware weighting under varying aggregation strengths $\alpha$. Higher $\alpha$ accelerates convergence but exacerbates representational skew by diminishing participation and degrading accuracy on high-staleness clients.}
\label{fig:async}
\vspace{-8pt}
\end{figure*}

We assess fairness in asynchronous FL by examining both client participation percentages (PP) and resulting local accuracy across devices under varying aggregation weights ($\alpha$). As shown in Fig.\ref{fig:async}, at $\alpha = 0.2$, high-end devices (\textit{HW$T_5$}, \textit{HW~$T_4$}) account for 62.1\% of updates, while mid-tier (\textit{HW~$T_3$}) and low-end devices (\textit{HW~$T_2$}, \textit{HW~$T_1$}) contribute 15.9\%, 13.7\%, and 8.2\%, respectively. At $\alpha = 0.6$, high-end participation rises to 80.2\%, while low-end devices drop to just 6.3\% and 4.5\%.

Importantly, this imbalance affects not only participation but also local performance. Low-end devices exhibit markedly lower accuracy—not only because they participate less frequently, but also because they often perform local training on outdated global models, making their updates less impactful. As $\alpha$ increases, this disparity widens: fast devices effectively steer the global model, while slower devices become marginalized in both influence and benefit.
These results emphasize that fairness in FL goes beyond balancing participation; it requires attention to equitable local outcomes. Addressing this challenge will require aggregation strategies that account for staleness and device diversity to improve both system-wide fairness and client-level utility.\\

\subsubsection{Privacy Loss under Synchronous and Asynchronous Aggregation}
\label{sec:privacy-analysis} 
Finally, we assess privacy loss under synchronous and asynchronous aggregation. Table~\ref{tab:privacy_loss} compares privacy loss ($\epsilon$) and per-device test accuracy degradation under LDP for staleness-aware \texttt{FedAsync} ($\alpha \in {0.2, 0.4, 0.6}$) and uniform \texttt{FedAvg}, across noise levels $\sigma \in {0.5, 1.0, 1.5, 2.0}$.

Under \texttt{FedAsync}, privacy loss is notably uneven across device tiers: high-end devices (e.g., \textit{HW~$T_5$}) accumulate substantially higher $\epsilon$ due to their frequent participation, exceeding $\epsilon > 35$ at $\sigma = 0.5$, while low-end devices (e.g., \textit{HW~$T_1$}) remain below $\epsilon < 10$ at the same noise level. This disparity intensifies at higher $\alpha$ values, where faster devices dominate the updates and thus incur greater privacy loss. In contrast, \texttt{FedAvg} enforces synchronized updates, resulting in uniform privacy loss across all devices.

A key observation under \texttt{FedAsync} is that although the same noise level is applied across all devices, high-end clients experience minimal test accuracy degradation (5.9\%–8.2\%), while low-end clients suffer severe drops (over 14\%). This stems from the fact that high-end devices, through frequent updates, exert disproportionate influence on the global model, making their local performance more resilient to noise. Conversely, low-end devices, affected by substantial staleness (e.g., $\tau = 7$ for \textit{HW~$T_1$}), contribute infrequent and down-weighted updates, making them more vulnerable to noise-induced degradation on their local test accuracy, especially as they often train on outdated global models. With \texttt{FedAvg}, all devices show almost similar accuracy degradation (6\%–11.5\%) due to equal participation and uniform noise application, providing fairness in both privacy and utility.

As noise increases, privacy loss ($\epsilon$) declines across all devices, confirming LDP’s protective effect, though at the cost of accuracy. Notably, setting $\alpha = 0.2$ offers a practical balance, moderating the dominance of high-end devices while preserving acceptable convergence speed, privacy, and fairness across the clients.\\

\begin{table*}[t]
\centering
\caption{Privacy loss ($\epsilon$) and test accuracy degradation across heterogeneous devices under varying noise levels, comparing \texttt{FedAsync} (Staleness-Aware, various $\alpha$) and \texttt{FedAvg}.}
\label{tab:privacy_loss}
\vspace{1mm}
\renewcommand{\arraystretch}{1}
\scalebox{0.75}{
\begin{tabular}{llcccccccc}
\toprule
\multirow{2}{*}{\textbf{Method}} & \multirow{2}{*}{\textbf{Device}} &
\multicolumn{2}{c}{\textbf{\(\sigma = 0.5\)}} &
\multicolumn{2}{c}{\textbf{\(\sigma = 1.0\)}} &
\multicolumn{2}{c}{\textbf{\(\sigma = 1.5\)}} &
\multicolumn{2}{c}{\textbf{\(\sigma = 2.0\)}} \\
\cmidrule(lr){3-4} \cmidrule(lr){5-6} \cmidrule(lr){7-8} \cmidrule(lr){9-10}
& & \textbf{\(\epsilon\)} & \textbf{Acc. Loss} & \(\epsilon\) & \textbf{Acc. Loss} &\textbf{\(\epsilon\)} & \textbf{Acc. Loss} & \textbf{\(\epsilon\)} & \textbf{Acc. Loss} \\
\midrule
\multirow{5}{*}{\shortstack{\textbf{FedAsync} \\ \textbf{(Staleness-Aware,} \\ \textbf{$\alpha=0.2$})}}
& \textit{HW~$T_1$} & \(\mathbf{12.57 \pm 0.35}\) & \(\mathbf{7.5 \pm 0.6}\) & \(\mathbf{2.02 \pm 0.05}\) & \(\mathbf{10.5 \pm 0.5}\) & \(\mathbf{0.56 \pm 0.04}\) & \(\mathbf{13.5 \pm 0.7}\) & \(\mathbf{0.21 \pm 0.02}\) & \(\mathbf{15.8 \pm 0.5}\) \\
& \textit{HW~$T_2$} & \(13.47 \pm 0.65\) & \(7.3 \pm 0.2\) & \(2.42\pm 0.03\) & \(10.1 \pm 0.5\) & \(0.83 \pm 0.02\) & \(13.2 \pm 0.5\) & \(0.37 \pm 0.01\) & \(15.5 \pm 0.8\) \\
& \textit{HW~$T_3$} & \(15.11 \pm 0.56\) & \(7.1 \pm 0.2\) & \( 3.20 \pm 0.05\) & \(9.3 \pm 0.3\) & \(1.11 \pm 0.03\) & \(12.6 \pm 0.5\) & \(0.48 \pm 0.05\) & \(14.8 \pm 0.6\) \\
& \textit{HW~$T_4$} & \(27.50 \pm 0.78\) & \(6.5 \pm 0.2\) & \(6.09 \pm 0.05\) & \(8.6 \pm 0.3\) & \(3.06 \pm 0.04\) & \(10.5 \pm 0.3\) & \(1.72 \pm 0.02\) & \(13.2 \pm 0.3\) \\
& \textit{HW~$T_5$} & \(\mathbf{31.75 \pm 0.69}\) & \(\mathbf{6.1 \pm 0.6}\) & \(\mathbf{6.21 \pm 0.03}\) & \(\mathbf{8.2 \pm 0.5}\) & \(\mathbf{3.33 \pm 0.05}\) & \(\mathbf{10.3 \pm 0.3}\) & \(\mathbf{1.98 \pm 0.06}\) & \(\mathbf{13.1 \pm 0.9}\) \\
\midrule
\multirow{5}{*}{\shortstack{\textbf{FedAsync} \\ \textbf{(Staleness-Aware,} \\ \textbf{$\alpha=0.4$})}}
& \textit{HW~$T_1$} & \(\mathbf{9.25 \pm 0.54}\) & \(\mathbf{7.6 \pm 0.4}\) & \(\mathbf{1.31 \pm 0.02}\) & \(\mathbf{11.3 \pm 0.4}\) & \(\mathbf{0.46 \pm 0.04}\) & \(\mathbf{13.9 \pm 0.2}\) & \(\mathbf{0.19 \pm 0.01}\) & \(\mathbf{15.9 \pm 0.6}\) \\
& \textit{HW~$T_2$} & \(10.51 \pm 0.55\) & \(7.4 \pm 0.3\) & \(1.52 \pm 0.01\) & \(10.9 \pm 0.7\) & \(0.77 \pm 0.06\) & \(13.8 \pm 0.4\) & \(0.26 \pm 0.05\) & \(15.7 \pm 0.8\) \\
& \textit{HW~$T_3$} & \(12.57 \pm 0.75\) & \(7.2\pm 0.6\) & \(2.02 \pm 0.04\) & \(9.9 \pm 0.6\) & \(1.01 \pm 0.08\) & \(12.8 \pm 0.3\) & \(0.38 \pm 0.05\) & \(15.0 \pm 0.6\) \\
& \textit{HW~$T_4$} & \(29.50 \pm 0.88\) & \(6.3 \pm 0.7\) & \(6.43 \pm 0.06\) & \(8.3 \pm 0.3\) & \(3.76 \pm 0.06\) & \(10.2 \pm 0.5\) & \(2.12 \pm 0.09\) & \(13.1 \pm 0.7\) \\
& \textit{HW~$T_5$} & \(\mathbf{33.58 \pm 0.65}\) & \(\mathbf{6.0 \pm 0.7}\) & \(\mathbf{6.50 \pm 0.03}\) & \(\mathbf{8.1 \pm 0.8}\) & \(\mathbf{3.87 \pm 0.02}\) & \(\mathbf{10.1 \pm 0.4}\) & \(\mathbf{2.78 \pm 0.04}\) & \(\mathbf{13.0 \pm 0.9}\) \\
\midrule
\multirow{5}{*}{\shortstack{\textbf{FedAsync} \\ \textbf{(Staleness-Aware,} \\ \textbf{$\alpha=0.6$})}}
& \textit{HW~$T_1$} & \(\mathbf{6.79 \pm 0.14}\) & \(\mathbf{7.8 \pm 0.6}\) & \(\mathbf{1.05 \pm 0.46}\) & \(\mathbf{11.9 \pm 0.5}\) & \(\mathbf{0.33 \pm 0.04}\) & \(\mathbf{14.1 \pm 0.4}\) & \(\mathbf{0.14 \pm 0.03}\) & \(\mathbf{16.1 \pm 0.8}\) \\
& \textit{HW~$T_2$} & \(8.30 \pm 0.24\) & \(7.5 \pm 0.5\) & \( 1.31 \pm 0.14\) & \(11.4 \pm 0.4\) & \(0.54 \pm 0.06\) & \(14.0 \pm 0.6\) & \(0.20 \pm 0.05\) & \(15.9 \pm 0.2\) \\
& \textit{HW~$T_3$} & \(10.60 \pm 0.11\) & \(7.3 \pm 0.6\) & \(2.35 \pm 0.24\) & \(10.2 \pm 0.4\) & \(0.91 \pm 0.08\) & \(12.9 \pm 0.5\) & \(0.31 \pm 0.05\) & \(15.2 \pm 0.1\) \\
& \textit{HW~$T_4$} & \(33.78 \pm 0.18\) & \(6.3 \pm 0.4\) & \(6.88 \pm 0.31\) & \(8.2 \pm 0.5\) & \(4.16 \pm 0.41\) & \(10.1 \pm 0.4\) & \(2.75 \pm 0.09\) & \(12.9 \pm 0.3\) \\
& \textit{HW~$T_5$} & \(\mathbf{35.12 \pm 0.49}\) & \(\mathbf{5.9 \pm 0.4}\) & \(\mathbf{7.50 \pm 0.33}\) & \(\mathbf{8.0 \pm 0.4}\) & \(\mathbf{4.78 \pm 0.15}\) & \(\mathbf{10.0 \pm 0.5}\) & \(\mathbf{3.63 \pm 0.01}\) & \(\mathbf{12.6 \pm 0.4}\) \\
\midrule

\textbf{FedAvg} & \textit{All Devices} & \(\mathbf{26.55 \pm 0.47}\) & \(\mathbf{6.0 \pm 0.3}\) & \(\mathbf{6.01 \pm 0.2}\) & \(\mathbf{7.9 \pm 0.3}\) & \(\mathbf{2.78 \pm 0.1}\) & \(\mathbf{9.5 \pm 0.3}\) & \(\mathbf{1.56 \pm 0.08}\) & \(\mathbf{11.5 \pm 0.3}\) \\
\bottomrule
\end{tabular}}
\end{table*}

\subsubsection{Privacy--Fairness--Efficiency Trade-off}
\label{eval:trade-off}
To summarize the evaluation and highlight the privacy--fairness--efficiency trade-offs, we find that while \texttt{FedAsync} achieves up to $9$–$10\times$ faster convergence to the 75\% target accuracy compared to \texttt{FedAvg}, this efficiency gain comes with significant disparities in both fairness and privacy. Specifically, high-end devices experience substantially higher privacy loss—up to $5$–$6\times$ more than low-end devices—posing potential risks in privacy-sensitive applications.

Under LDP, low-end devices show greater accuracy degradation (up to 16\%) due to their infrequent and stale updates, which are more vulnerable to the noise injected for privacy protection. Notably, the global model is effectively shaped by the high-end devices, which dominate the update stream; as a result, low-end devices experience greater variance and reduced alignment between the global model and their local data, further marginalizing their contributions and amplifying both fairness and privacy gaps.

Reducing the aggregation weight ($\alpha$) can partially mitigate these disparities, improving participation fairness and reducing the privacy gap, though at the cost of slower convergence. Overall, while \texttt{FedAsync} offers fast, scalable training, it introduces unequal privacy risks and fairness concerns, particularly for resource-constrained clients in privacy-preserving settings. These findings underscore the need to jointly tune aggregation strategies, staleness handling, and privacy mechanisms to achieve a balanced, application-specific trade-off across efficiency, fairness, and privacy.

\section{Discussion and Key Takeaways}
\label{sec:discussion}
Our empirical findings highlight how device heterogeneity—ubiquitous in real-world FL deployments—introduces fundamental challenges across efficiency, fairness, and privacy. While asynchronous methods like \texttt{FedAsync} deliver remarkable efficiency gains, achieving up to 9–10$\times$ faster convergence over \texttt{FedAvg} by bypassing stragglers and prioritizing high-end devices, these gains come with hidden costs.

\texttt{FedAsync} accelerates learning by letting fast devices dominate updates, especially under staleness-aware aggregation, which improves stability by down-weighting outdated contributions. However, this disproportionately amplifies the role of high-end devices, with their participation share exceeding 80\% at higher $\alpha$ values, while low-end devices fall below 5\%. Even under IID data, this imbalance skews the global model and leads low-end clients to train on increasingly outdated global models, degrading their local accuracy and marginalizing their influence.

The privacy picture is perhaps even more concerning: high-end devices (e.g., \textit{HW~$T_5$}) accumulate privacy loss exceeding $\epsilon > 35$ at $\sigma = 0.5$, while low-end devices (e.g., \textit{HW~$T_1$}) remain below $\epsilon < 10$. This sharp disparity poses serious challenges, especially in sensitive domains like healthcare or finance, where unequal privacy guarantees can have critical implications. Raising the LDP noise level helps narrow these privacy gaps by lowering overall privacy loss, but it also degrades both global and local accuracy, underscoring a delicate trade-off between fairness, privacy, and utility. Moreover, uniform LDP disproportionately impacts low-end devices, reducing their local accuracy by up to 16\% because their marginal, infrequent updates—often trained on stale global models—are more vulnerable to added noise, further amplifying performance disparities.

Real-world FL deployments must account for domain-specific constraints. While fairness and privacy are paramount in healthcare applications, low-latency FL strategies are prioritized in automotive systems. Future research should explore the integration of adaptive privacy mechanisms with fairness-aware aggregation to develop scalable, privacy-aware, and equitable FL systems suited for diverse, resource-constrained environments.
    
\vspace{0.5em}
\noindent\textbf{Future Directions.} Building on these insights, we outline three promising directions to advance the design of FL systems under real-world heterogeneity:

\begin{itemize}
    \item \textbf{Joint Aggregation--Privacy Adaptation.} 
    Future FL protocols should treat aggregation and privacy as interdependent, not separate, challenges. Dynamically adjusting both client weights and privacy budgets (e.g. in LDP) based on real-time signals—such as participation frequency, staleness, and device capabilities—could help achieve more equitable privacy guarantees without sacrificing convergence. Extending adaptive DP frameworks (e.g.,~\cite{liu2021projected}) into asynchronous settings represents a compelling step forward.

    \item \textbf{Fairness-Aware Privacy Calibration.} Current LDP approaches often apply uniform noise across clients, ignoring the natural disparities in participation frequency and staleness that arise in heterogeneous systems. A promising direction is to develop client-specific privacy calibration strategies that dynamically adjust noise levels based on each client’s update frequency, staleness, and influence on the global model. For example, frequently participating high-end devices could tolerate stronger privacy noise to limit their dominant influence, while low-end devices with infrequent updates could operate under lighter noise to preserve local utility. This would help balance fairness and privacy across diverse clients. Recent work on fairness-aware aggregation~\cite{zhu2022online} offers a useful foundation for this direction.

    \item \textbf{Personalized FL with Privacy Guarantees.} Inspired by multi-task and personalized FL frameworks~\cite{deng2020adaptive}, future systems should support models with client-personalized components tailored to device capabilities and privacy needs. This would allow low-end clients to achieve meaningful utility even under stronger privacy constraints, reducing their dependence on frequent global updates and improving system inclusivity.
\end{itemize}

\vspace{0.5em}
\noindent In sum, advancing FL under heterogeneity requires rethinking aggregation, privacy, and fairness as tightly coupled objectives. Designing adaptive, privacy-aware, and client-personalized FL protocols offers a promising path toward democratizing access to FL without sacrificing system-wide efficiency.

\section{Conclusion}
This work investigates the hidden privacy costs of asynchronous FL under real-world device heterogeneity, revealing fundamental trade-offs between efficiency, fairness, and privacy. While \texttt{FedAsync} accelerates convergence and reduces training time, it disproportionately favors high-end devices, exacerbating fairness disparities. Our analysis shows that high-end clients experience up to \textit{6×} greater privacy loss due to frequent updates, while uniform noise scaling fails to deliver equitable privacy protection, degrading the accuracy of low-end clients whose updates are both infrequent and stale. These findings highlight a key limitation of current FL aggregation methods, which often prioritize efficiency and fairness while overlooking privacy risks shaped by device heterogeneity.
Addressing this challenge will require adaptive FL frameworks that jointly calibrate aggregation and privacy mechanisms in response to client participation patterns and staleness profiles. Such designs are essential to enable scalable, equitable, and privacy-preserving learning across real-world edge deployments. Ultimately, bridging the gap between efficiency, fairness, and privacy is not just a technical goal—it is a prerequisite for FL to fulfill its promise of inclusive, ethical, and globally deployable machine intelligence.

\section{Acknowledge}
This work has been supported by the H2020 ECSEL EU project Distributed Artificial Intelligent System (DAIS) under grant agreement No. 101007273, and the Knowledge Foundation within the framework of INDTECH (Grant No. 20200132) and INDTECH+ Research School project (Grant No. 20220132). Additional funding was provided by the Research Council of Norway (RCN), Romania UEFISCDI, Spain CSCJA, Sweden Forte, under the Transforming Health and Care Systems (THCS, GA No. 101095654, EU Horizon Europe).

\bibliographystyle{unsrt}
% \bibliography{ref}

% \input{main.bbl}

\end{document}